\documentclass[12pt]{article}
\pdfoutput=1
\usepackage{graphicx,psfrag,epsf,color}
\usepackage[dvipsnames]{xcolor}
\usepackage{amsmath,amssymb,amsfonts, booktabs}
\usepackage{appendix}
\usepackage{multirow}
\usepackage{geometry}
\usepackage{float}
\usepackage{mathrsfs}  
\usepackage{authblk}
\usepackage{array}
\usepackage{cite}
\usepackage{slashed}
\usepackage{graphicx}
\graphicspath{ {Images/} }
\usepackage{rotating}
\usepackage{slashed, cancel}
\usepackage[caption=false]{subfig}
\usepackage[compat=1.1.0]{tikz-feynman} 
\usepackage{hyperref}
\newcounter{MBQ}

\usepackage{tabularx}
\newcolumntype{C}{>{\centering\arraybackslash}X}

\newcommand{\eps}{\epsilon}

\def\be{\begin{equation}}
\def\ee{\end{equation}}
\def\beq{\begin{eqnarray}}
\def\eeq{\end{eqnarray}}
\newcommand{\bea}{\begin{eqnarray}}
\newcommand{\eea}{\end{eqnarray}}
\newcommand{\beas}{\begin{eqnarray*}}
\newcommand{\eeas}{\end{eqnarray*}}

\newcommand{\alem}{\alpha_{\rm em}}

\newcommand{\bra}[1]{\big\langle{#1}\big\vert}
\newcommand{\ket}[1]{\big\vert{#1}\big\rangle}

\begin{document}
\allowdisplaybreaks

\begin{titlepage}

\begin{flushright}
{\small
TUM-HEP-1349/21\\
Nikhef-2021-015\\
July 8, 2021 \\
}
\end{flushright}

\vskip1cm
\begin{center}
{\Large \bf\boldmath QED factorization of two-body non-leptonic and semi-leptonic $B$ to charm decays }
\end{center}

\vspace{0.5cm}
\begin{center}
{\sc Martin~Beneke,$^a$ Philipp B\"oer,$^a$ 
Gael Finauri,$^{a}$ K. Keri Vos$^{b,c}$} \\[6mm]
{\it $^a$Physik Department T31,\\
James-Franck-Stra\ss{}e~1, 
Technische Universit\"at M\"unchen,\\
D--85748 Garching, Germany\\[0.3cm]

{\it $^b$Gravitational 
Waves and Fundamental Physics (GWFP),\\ 
Maastricht University, Duboisdomein 30,\\ 
NL-6229 GT Maastricht, the
Netherlands}\\[0.3cm]

{\it $^c$Nikhef, Science Park 105,\\ 
NL-1098 XG Amsterdam, the Netherlands}}
\end{center}

\vspace{0.6cm}
\begin{abstract}
\vskip0.2cm\noindent
The QCD$\times$QED factorization is studied for two-body non-leptonic and semi-leptonic $B$ decays with heavy-light final states. These non-leptonic decays, like $\bar{B}^0_{(s)}\to D^+_{(s)} \pi^-$ and $\bar{B}_d^0 \to D^+ K^-$, are among the theoretically cleanest non-leptonic decays as penguin loops do not contribute and colour-suppressed tree amplitudes are suppressed in the heavy-quark limit or even completely absent. Advancing the theoretical calculations of such decays requires therefore also a careful analysis of QED effects. Including QED effects does not alter the general structure of factorization which is analogous for both semi-leptonic and non-leptonic decays. For the latter, we express our result as a correction of the tree amplitude coefficient $a_1$. At the amplitude level, we find QED effects at the sub-percent level, which is of the same order as the QCD uncertainty. We discuss the phenomenological implications of adding QED effects in light of discrepancies observed between theory and experimental data, for ratios of non-leptonic over semi-leptonic decay rates. At the level of the rate, ultrasoft photon effects can produce a correction up to a few percent, requiring a careful treatment of such effects in the experimental analyses.  
\end{abstract}
\end{titlepage}



\section{Introduction}

Recently, the QCD factorization formula for non-leptonic $B$ decays into two light mesons ~\cite{Beneke:1999br,Beneke:2000ry} has been generalized to include QED~\cite{Beneke:2020vnb}. In this paper, we discuss the implications of QED on the factorization of non-leptonic heavy-to-heavy decays, specifically $B \to D^{(*)} L$ where $L$ is a light meson $L= \pi, K^{(*)}, \rho$. Schematically, in QCD \cite{Beneke:2000ry}
\begin{equation}
 \left\langle D^{(*)} L | Q_i |\bar{B} \right\rangle = 
F^{B \to D^{(*)}} \!\times T_i * \phi_{L} \ ,
\label{eq:QCDF}
\end{equation} 
which is valid in the heavy-quark limit applied to both bottom and charm mesons. The factorization formula describes the matrix element of operator $Q_i$ in terms of the heavy-to-heavy transition form factor $F^{B\to D^{(*)}}$, the light-cone distribution amplitude (LCDA) $\phi_L$ of the light-meson $L$ and their convolution with the perturbatively calculable scattering kernel $T_i$. Contrary to $B$ decays into two light mesons, here only the ``form factor" term contributes, while ``spectator scattering" is power suppressed in the $\Lambda_{\rm QCD}/m_Q$ counting and thus absent at leading order, where $m_Q$ is the heavy quark mass. 

The QCD corrections to the short-distance kernels 
$T_i$ are already known to 
$\mathcal{O}(\alpha_s^2)$ (NNLO)
\cite{Huber:2016xod}. Recently, these heavy-to-heavy decays have received renewed attention, due to deviations between theoretical predictions and experimental measurements of $\bar{B}_d^0 \to D^+ K^-$ and $\bar{B}^0_{(s)} \to D_{(s)}^+ \pi^-$ \cite{Bordone:2020gao, Cai:2021mlt, Bordone:2021cca}. Such decays are dominated by colour-allowed tree topologies described by the topological tree-amplitude $a_1(D^+ L^-)$. Taking as a specific example $\bar B_d^0 \to D^+ K^-$, the amplitude becomes \cite{Beneke:2000ry}:
\begin{align}
 {\mathcal{A}}(\bar B_d^0 \to D^+ K^-) & = i \, \frac{G_F}{\sqrt{2}} \, V^*_{us} \, V_{cb} \; a_1(D^+ K^-) \, f_K  \, (m_B^2-m_D^2)\, F_0^{B\to D}(m^2_K) \, , \label{eq:defa1}
\end{align}
where $f_K$ is the kaon decay constant. The coefficient $a_1(D^+K^-)$ is known up to NNLO \cite{Huber:2016xod}. Power corrections of $\mathcal{O}(\Lambda_{\rm QCD}/m_Q)$ were discussed in \cite{Beneke:2000ry} and were estimated to be at the subpercent level at most in \cite{Bordone:2020gao} using sum-rule techniques. Not included in these estimates are QED effects, which are expected to be smaller. 

In this paper, we discuss the impact of QED on heavy-to-heavy non-leptonic decays extending our work in \cite{Beneke:2020vnb}. Once including QED effects, the branching fraction and amplitudes are no longer infrared-finite, and the process that should be considered is $B\to D^{(*)} L (\gamma)$  where 
$\gamma$ is any number of soft photons with total 
energy less than $\Delta E$ in the $B$ meson rest frame. At scales above $\Delta E$, QED corrections are purely virtual. The factorization of the so-called ``non-radiative'' amplitude along the lines of~\cite{Beneke:2020vnb} can then be formulated if $\Delta E\ll \Lambda_{\rm QCD}$. We calculate these virtual corrections to non-radiative $\bar{B}^0\to D^{+(*)} L^-$ decays and combine them with ultrasoft radiation to obtain the infrared finite physical branching fractions. 

The paper is organized as follows: First, we give the QED generalized factorization formula and the explicit results for the hard scattering kernels. We then discuss the ultrasoft effects and the phenomenological implications of QED on Standard Model predictions.

\section{Factorization Formula}
In this work we consider the decay of a $\bar{B}_{(s)}$ meson into a heavy $D^{(*)+}_{(s)}$ meson and a light meson $L^-$ ($L=\pi, \, \rho,  \, K, \, K^\ast)$ mediated by the current-current operators for $b\to c$ transitions, 
given by the weak Hamiltonian
\begin{equation}             
\mathcal{H}_{\rm eff} = \frac{G_F}{\sqrt{2}} \,V_{uD}^* V_{cb}
           \left( C_1 Q_1 + C_2 Q_2\right) + \mathrm{h.c.}
\end{equation}
with the CMM operator basis~\cite{Chetyrkin:1997gb}  
\begin{align}
        Q_1 &= [\bar c \gamma^\mu T^a (1 -\gamma_5) b]
                [\bar D \gamma_\mu T^a (1 - \gamma_5) u] ,
\nonumber\\
        Q_2 &= [\bar c \gamma^\mu (1 -\gamma_5) b]
                 [\bar D \gamma_\mu (1 - \gamma_5) u],
\label{eq:weakham}
\end{align}
and $D=d$ or $s$. $T^a$ denotes the SU(3) colour generator in the fundamental representation. In general, the $B$ meson decays to $D^{(*)} L$ via (a combination of) quark topologies, corresponding to colour-allowed tree decay (class-I), colour-suppressed tree decay (class-II) and via weak annihilation. QCD and electroweak penguin topologies do not contribute. Both, the colour-suppressed tree and weak annihilation topologies are power-suppressed with respect to the colour-allowed tree topology \cite{Beneke:2000ry}. In the following, we consider only $\bar{B}_{(s)}^0 \to D_{(s)}^{(*)+} L^-$ decays which proceed (predominantly) through the colour-allowed tree topology. As in the derivation of the QCD factorization formula \cite{Beneke:2000ry}, we adopt the power counting $z\equiv m_c^2/m_b^2 \sim \mathcal{O}(1)$. 

Similar as for $B$ decays into two light mesons, the QCD factorization formula can be extended to include QED as 
\begin{eqnarray}
\label{eq:QEDF}
\left\langle D^+ L^- | Q_i |\bar B\right\rangle &=& 4i E_D E_L \,
\zeta^{BD}_{Q_L}(m_L^2)
\, \int_0^1 du \, 
H_{i,Q_{L}}(u,z)
\, f_{L} \Phi_{L}(u) 
 \, , 
\end{eqnarray}
where here $f_L$ is the renormalization-scale independent QCD decay constant.\footnote{This definition differs slightly from \cite{Beneke:2020vnb} where a QED generalized decay constant was introduced. Here all QED effects are put into the definition of the LCDA $\Phi_L$ as discussed in detail in \cite{BLCDApaper}.} The kernels, form factors and LCDAs depend on the electric charge $Q_L$ of the light meson $L$.  These objects are generalized to include virtual short- and long-distance photon exchange. As given in \eqref{eq:QEDF}, these objects are single-scale quantities. We shall later replace the HQET form factor $\zeta^{BD}_{Q_L}$ by a suitable physical quantity, making use of the fact that an analogous factorization theorem holds for the non-radiative semi-leptonic $\bar{B} \to D ^+\ell^- \bar{\nu}_\ell$ amplitude at the kinematical point $q^2=m_L^2$ and $E_\ell = m_b(1-z)/2$ (as discussed in Sec.~\ref{sec:semileptonic} and \cite{Beneke:2020vnb}). This procedure is akin to using the full QCD rather than HQET $B\to D$ form factor in QCD alone.
As stated above, we only consider $L^-$, with $Q_L = -1$. In the following, we therefore omit the $Q_L$ subscript. In \eqref{eq:QEDF}, we choose a specific normalization for the form factor, where $E_D= (m_B^2+m_D^2-m_L^2)/(2 m_B)$ and $E_L= (m_B^2-m_D^2+m_L^2)/(2 m_B)$ are the energies of the $D$ and $L$ mesons in the $B$ rest frame, respectively. As we treat the charm quark as heavy, the scattering kernels $H_i$ depend on $z\equiv m_c^2/m_b^2$ as well as on the momentum fraction $u$. Finally, we note that the factorization formula also holds for $\bar{B}\to D^{*+} L^-$ and $\Lambda_b \to \Lambda_c^+ L^-$, as was the case without QED.

\begin{figure}[t]
\centerline{\includegraphics[scale=0.8]{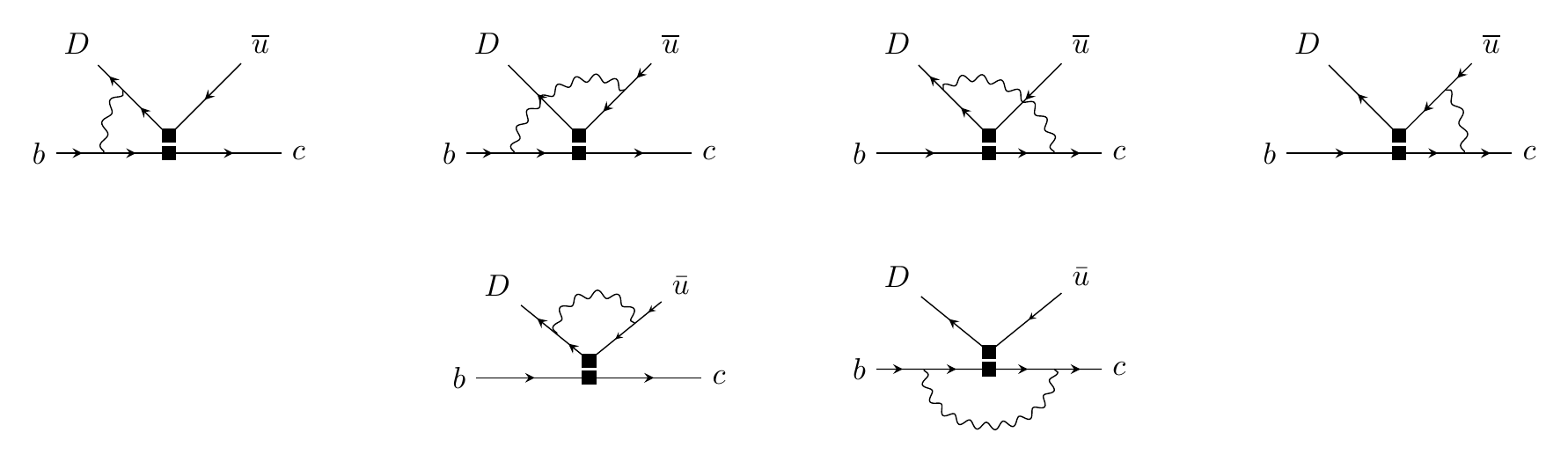}}
\caption{\label{fig:dia} One-loop $\alem$-contributions to $b\to c \bar{u}D$ transitions of the operator $Q_2$. }
\end{figure}


\section{Details on the calculation}

To derive the factorization formula already given in \eqref{eq:QEDF}, we follow the procedure discussed in detail in \cite{Beneke:2020vnb} for $B$ decays into two light mesons. There all the relevant QED generalized operator definitions were given. The heavy-to-heavy calculation in this paper is in fact simpler than that of the $B$ into two light mesons as now only QCD$\times$QED $\to$ SCET$_{\rm I}$ matching is required. In the following, we compute the $\mathcal{O}(\alem \alpha_s^0)$ contributions to the scattering kernel $H_i$. The corresponding quark-level diagrams are shown in Fig.~\ref{fig:dia}. 

We proceed by matching the QCD operators $Q_{1,2}$ onto the QED generalized SCET$_{\rm I}$ operators
\begin{equation}
 \mathcal{O}_\mp(t)= \bar{\chi}^{(D)}(tn_-) \frac{ \slashed{n}_{-}}{2} (1-\gamma_5) \chi^{(u)}(0) \, \bar{h}_{v'}(0) \slashed{n}_{+} (1\mp\gamma_5)S_{n_+}^{\dagger (Q_{L})}(0) h_v(0) \, , \label{eq:Op1}
\end{equation}
where the bottom (charm) quark is treated as a static quark $h_v \; (h_{v'})$ with velocity $v \;(v')$. Here $n_+^\mu$ is a light-like reference vector defined along the direction of motion of the light meson $L$, and $n_-^\mu$ refers to the opposite direction such that $n_+^2=n_-^2=0$ and $n_+n_-=2$. The difference with QCD is the presence of the soft QED Wilson line that depends on the electric charge of the light meson $L$ \cite{Beneke:2020vnb}:
\begin{eqnarray}
\label{eq:softwilsonlines}
S^{(Q_L)}_{n_\pm}(x) &=& \exp \left\{ -i Q_L e \int_0^{\infty} \!ds \,
 n_\pm A_{s}(x + s n_\pm) \right\} 
\nonumber
\end{eqnarray}
The QCD part of the soft Wilson line cancels in \eqref{eq:Op1}, because the operators are colour singlets. 
In pure QED, to all orders in $\alem$, 
\begin{equation}
    \left\langle D^{(*)} L | Q_1 |\bar B\right\rangle \equiv \langle Q_1 \rangle =0 \ ,
\end{equation}
because the colour-octet QCD operator $Q_1$ cannot match onto the colour-singlet SCET operators $\mathcal{O}_\mp$. 
The renormalized matrix element of $Q_2$ is matched as\footnote{The hard-scattering kernels are functions of $t$ and their product with $\mathcal{O}_\mp$ is in fact a convolution.}
\begin{equation}
\label{eq:Q2matel}
\langle Q_2 \rangle = H_- \langle \mathcal{O}_- \rangle + H_+ \langle \mathcal{O}_+ \rangle \ .
\end{equation}
The matching coefficients $H_{+}$ and $ H_{-}$ can be expressed as
\begin{align}
\label{eq:oneloopmaster}
H_- &= A_{2-}^{(0)} + \frac{\alem}{4\pi}\left\{ A_{2-}^{(1)} + Z_\textup{ext}^{(1)} - Y^{(1)} + Z_{2j}^{(1)} A_{j-}^{(0)}
\right\} \,, \\
H_+ &=\frac{\alem}{4\pi} A_{2+}^{(1)} \ ,
\end{align}
where the superscript indicates the expansion coefficients in powers of $\alem/(4\pi)$. Here $A_{2-}^{(0)}\;(A_{2\mp}^{(1)})$ are the bare tree-level (one-loop) on-shell matrix elements of operator $Q_2$. We note that $H_+=0$ if $m_c\to 0$, because the chirality flipped operator does not contribute for light final states. The one-loop QED operator renormalization constants are given by $Z_{ij}$ in (16) of \cite{Beneke:2020vnb}, where $j$ also runs over evanescent operators \cite{Beneke:2009ek} (see also \cite{Huber:2016xod}).  The external field renormalization contribution $Z_\textup{ext}^{(1)}$ accounting for the one-loop on-shell renormalization of the $b$-quark and $c$-quark fields is
\begin{equation}
\label{eq:Zext}
Z_\textup{ext}^{(1)} = -Q_d^2\biggl[\frac{3}{2\epsilon} + \frac{3}{2}L_b + 2 \biggr] - Q_u^2\biggl[\frac{3}{2\epsilon} + \frac{3}{2}L_b - \frac{3}{2}\ln z + 2 \biggr] \, ,
\end{equation}
where 
\begin{equation}
    L_b \equiv \ln \frac{\mu^2}{m_b^2} \ .
    \end{equation}

\begin{figure}[t]
\centerline{\includegraphics[scale=1.2]{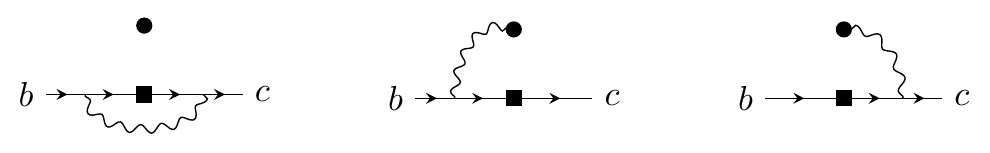}}
\caption{\label{fig:Zhh} $\alem$-corrections to the heavy-to-heavy current. Here the dot represents the soft Wilson line $S_{n_+}$.}
\end{figure}

Similar to \cite{Beneke:2020vnb}, we write the SCET operator renormalization factor $Y$ in two parts:
\begin{equation}
\label{eq:Y1}
Y^{(1)}(u,v) = Z_{hh}^{(1)} \delta (u-v) + Z^{(1)}_{\bar{C}}(u,v)
\end{equation}
where $Z^{(1)}_{\bar{C}}(u,v)$ is the anti-collinear kernel computed in (22) in \cite{Beneke:2020vnb}. The generalized heavy-to-heavy current is now defined as
\begin{equation}
\bar{h}_{v'}(0) \slashed{n}_+ (1-\gamma_5) S_{n_+}^{\dagger (Q_L)}(0) h_v(0)
 ,
 \end{equation}
where the soft Wilson line $S_{n_+}$ arises from the soft decoupling of the anti-collinear fields. We obtain $Z_{hh}$ at one-loop by calculating the diagrams in Fig.~\ref{fig:Zhh} by regularizing the infrared (IR) divergences with an off-shellness $k_q^2$ for the $q=u,d$ quarks. As discussed in Appendix A of~\cite{Beneke:2019slt} (see also \cite{Beneke:2020vnb}), this also requires modifying the soft Wilson line propagator.
In the $\overline{\rm MS}$ scheme, we obtain
\begin{equation}
\begin{split}
\label{eq:Zhh}
Z_{hh}^{(1)} &= \frac{1}{\epsilon}\biggl\{ -Q_d^2\biggl(\frac{1+z}{1-z}\ln z + 2 \biggr) +2 Q_{L} Q_d \biggl(\frac{\ln z}{1-z} +i \pi + 1\biggr) \\ 
&+ Q_{L}^2 \biggl(\frac{1}{\epsilon} + \ln \frac{\mu^2}{z \delta_{\bar{c}}^2}-1 \biggr)
\biggr\} \ ,
\end{split}
\end{equation}
where $\delta_{\bar c}\equiv k_{q}^2/(n_-k_{q})$ is the off-shell regulator in the soft Wilson line.   
For neutral $L$, this reduces to the QCD result (see (39) in~\cite{Huber:2016xod}) after replacing the charge factors $Q_d^2\to C_F, Q_L \to 0$. Finally, $Y^{(1)}$ can be obtained using \eqref{eq:Y1} by adding \eqref{eq:Zhh} and $Z_{\bar{C}}$ from (22) in \cite{Beneke:2020vnb}. It is independent of the IR regulator $\delta_{\bar{c}}$ as it should be.

\subsection{Hard matching coefficients}\label{sec:hmc}
At tree level, only $A_{2-}^{(0)} = 1$ contributes. At one-loop, we obtain the hard-scattering kernels by computing the on-shell matrix elements $A^{(1)}_{2\mp}$ from the diagrams in Fig.~\ref{fig:dia}. The hard-scattering kernels are obtained using the renormalization factors in \eqref{eq:oneloopmaster}. The $1/\epsilon$ poles properly cancel, and we find
\begin{align}
\label{eq:H2}
H_-^{(1)} =& -Q_d^2 \biggl\{\frac{L_b^2}{2} + L_b\biggl(\frac{5}{2} -2\ln(u(1-z)) \biggr) + h\bigl(u(1-z)\bigr) + \frac{\pi^2}{12} + 7 \biggr\} \nonumber \\ 
&-Q_u^2 \biggl\{\frac{L_c^2}{2} + L_c\biggl(\frac{5}{2}+2\pi i -2\ln \Bigl(\bar{u}\frac{1-z}{z}\Bigr) \biggr) + h\Bigl(\bar{u}\Bigl(1-\frac{1}{z}\Bigr)\Bigr) + \frac{\pi^2}{12} + 7   \biggr\} \nonumber \\
&+Q_d Q_u \biggl\{\frac{L_b^2}{2} + \frac{L_c^2}{2} -6 L_\nu +2L_b\biggl(2-\ln(\bar{u}(1-z))\biggr) \nonumber \\
&-2L_c\biggl(1-i\pi+\ln\Bigl(u\frac{1-z}{z}\Bigr) \biggr) +g\bigl(\bar{u}(1-z)\bigr) + g\Bigl(u\Bigl(1-\frac{1}{z}\Bigr) \Bigr) +\frac{\pi^2}{6} -12 \biggr\} \nonumber \\
&+Q_d Q_u f(z)\,, \\
H_+^{(1)} = &-Q_d^2 \sqrt{z} \, w\bigl(u(1-z)\bigr) -Q_u^2 \frac{1}{\sqrt{z}} \;w\Bigl(\bar{u}\Bigl(1-\frac{1}{z}\Bigr)\Bigr) -Q_d Q_u \sqrt{z}\,\frac{\ln z}{1-z} \,,
\end{align}
where we have defined 
\begin{equation}
z\equiv \frac{m_c^2}{m_b^2} \ , \qquad L_c \equiv \ln \frac{\mu^2}{m_c^2} = L_b - \ln z \,, \qquad        L_\nu \equiv \ln \frac{\nu^2}{m_b^2} \ ,
\end{equation}
where $\nu$ refers to the scale of the Wilson coefficient $C_2(\nu)$. In addition, we defined the functions
\begin{equation}
\begin{split}
h(s) &\equiv \ln^2 s -2\ln s +\frac{s \ln s}{1-s}-2 \text{Li}_2\Bigl(\frac{s-1}{s}\Bigr) \,, \\
g(s) &\equiv h(s) -\frac{3 s \ln s}{1-s} \,, \\
f(z) &\equiv \biggl(1-\frac{1+z}{1-z}\ln z \biggr)L_b + \frac{\ln z}{1-z}\biggl( \frac{1}{2}(1+z)\ln z-2-z \biggr) \,, \\
w(s) &\equiv \frac{1-s+s\ln s}{(1-s)^2} \,.
\end{split}
\end{equation}
To extract the imaginary parts from these functions $z$ is understood as $z-i \epsilon$. The function $f(z)$ contains the entire contribution of the last diagram of Fig.~\ref{fig:dia}. For the other contributions in $H_-^{(1)}$ in \eqref{eq:H2}, we note the symmetry when exchanging the contributions involving the charm and bottom quarks, as they are both treated as heavy quarks. Explicitly, the $Q_u^2$ term, arising from the fourth diagram in Fig.~\ref{fig:dia}, is obtained from the $Q_d^2$ term (first diagram) when switching $L_b\to L_c, Q_d\to Q_u, u\to \bar{u}$ and $z\to 1/z$. The second and third diagram in Fig.~\ref{fig:dia} give the $Q_u Q_d$ terms, which are symmetric under the exchange of $z\to 1/z, u\to \bar{u}$ and $L_b\to L_c$. 
We emphasize that the hard-scattering kernel $H_-$ in \eqref{eq:H2} diverges when taking the $z\to 0$ limit: 
\begin{align}
H_-^{(1)}(z \to 0) &= \frac{Q_u^2}{2} \biggl( \ln^2 z - (2 L_b +1)\ln z \biggr) + \rm{finite~terms}  \ , \\
H_+^{(1)}(z \to 0) &= 0 \,.
\end{align}
In pure QCD, this limit can be smoothly taken as the collinear divergences between the diagrams where the gluon couples to the charm quark cancel, and the only divergences remain in the factorizable diagrams (given in the second line in Fig.~\ref{fig:dia}). In QCD, these terms are then absorbed into the form factor. In QED, the collinear divergences will be cancelled when normalizing to the semi-leptonic decay as discussed below (see also \cite{Beneke:2020vnb}). 

Taking now the matrix element of \eqref{eq:Q2matel}, requires the QED generalized light-meson LCDA $\Phi_L$ introduced in (51) of \cite{Beneke:2020vnb} and further discussed in \cite{BLCDApaper}. The matrix element is normalized by $E_L= (m_B^2-m_D^2+m_L^2)/(2 m_B)$ and is multiplied by the soft rearrangement factor $R_{\bar{c}}^{(Q_L)}$. This rearrangement, discussed in detail in Sec. 4.1 of \cite{Beneke:2020vnb}, removes the overlap of the soft and collinear sectors and ensures the renormalizability of the light-meson LCDA.  In addition, we need to introduce QED generalized HQET form factors $\zeta^{BD}$ and $\zeta^{BD^*}$ which we define similar to (52) of \cite{Beneke:2020vnb} by replacing the $\chi_{C}\to h_{v'}$:
\begin{align}
\label{eq:SCET1FFs}
\langle  D|\frac{1}{R_{\bar{c}}^{(Q_L)}} \bar{h}_{v'}(0) \slashed{n}_+ S_{n_+}^{\dagger(Q_L)}(0) h_v (0)|\bar{B}\rangle  =   4 E_{D}\, \zeta^{B D}_{Q_L}\ , 
\end{align}
where our normalization differs from the standard HQET Isgur-Wise function. For the $D^*$, we define
\begin{align}
\label{eq:SCET1FFs2}
\langle  D^{*}|\frac{1}{R_{\bar{c}}^{(Q_L)}} \bar{h}_{v'}(0) \slashed{n}_+ \gamma_5 S_{n_+}^{\dagger(Q_L)}(0) h_v (0)|\bar{B}\rangle  = - 4 E_{D^{*}} \; \epsilon^{*} \cdot v\, \zeta^{B D^{*}}_{ Q_L}\ , 
\end{align}
where $\epsilon^*$ is the $D^*$ polarization vector and $E_{D^{(*)}} = (m_B^2+m_{D^{(*)}}^2-m_L^2)/(2 m_B)$.
Due to the Wilson line $S_{n_+}^{(Q_L)}$ and due to the soft-subtraction term $R_{\bar{c}}^{(Q_L)}$, the form factor $\zeta$ depends on the charge of the light meson. In the following, as we only consider charged $L^-$, we omit this additional subscript. Defining then $H\equiv H_-+H_+$ and $H^* \equiv H_- - H_+$, allows us to write the factorization formula, already quoted in \eqref{eq:QEDF}, as
\begin{align}
\langle D^{+} L^-| Q_2 |\bar{B} \rangle &= 4i E_{D} E_{L} \zeta^{BD}\int_0^1 du \;  H(u,z) f_L \Phi_L(u) \ , \label{eq:factscet1}\\
\langle D^{*+} L^-| Q_2 |\bar{B} \rangle &= 4i E_{D^*} E_{L}\;\epsilon^*\cdot v\;\zeta^{BD^{*}}\int_0^1 du\; H^*(u,z) f_L \Phi_L(u) \ .  
\label{eq:factscet}
\end{align}
Finally, we note that the above holds also for longitudinally polarized vector mesons ($L=\rho, K^*$), while decays to transversely polarized states are power suppressed.

\subsection{Semi-leptonic decay}
\label{sec:semileptonic}
\subsubsection{Factorization formula}
The QCD$\times$QED factorization formulas for the non-radiative semi-leptonic $\bar B \to D^{+(*)} \ell^- \bar{\nu}_{\ell}$ amplitude are very similar to those of non-leptonic $\bar{B}\to D^{+(*)}L^-$ decays in \eqref{eq:factscet1} and \eqref{eq:factscet}. We define
\begin{align}
\mathcal{A}^{{\rm sl},D^{(*)}}_{\text{non-rad}}& = 
\frac{G_F}{\sqrt{2}} V_{cb} C_{\rm sl}\,
\bra{D^{+(*)} \ell^- \bar{\nu}_\ell}Q_{\rm sl} \ket{\bar B} \ , 
\end{align}
where the semi-leptonic operator is $Q_{\rm sl} = \bar{c}\gamma^\mu (1-\gamma_5)b \, \bar{\ell}\gamma_\mu(1-\gamma_5)\nu$. $C_{\rm sl}$ is the semi-leptonic hard Wilson coefficient. Following \cite{Beneke:2020vnb}, the QCD$\times$QED $\to$ SCET matching of the amplitude gives 
\begin{equation}\label{eq:QEDFsemi}
\langle D^+ \ell^- \bar{\nu}_\ell| Q_{\rm sl} |\bar{B} \rangle = 4 i E_{D} E_{\rm sl} \, Z_{\ell} \,\zeta^{BD}(E_\ell,q^2) H_{\rm sl}(E_\ell,q^2,z) \ ,
\end{equation}
where we defined the spinor product
\begin{equation}
i E_{\rm sl} \equiv \bar{u}(p_\ell) 
\,\frac{\slashed{n}_-}{2} (1-\gamma_5) v_{\nu_{\ell}}(p_{\nu_\ell}) \ .
\end{equation} 
$Z_{\ell}$ denotes the renormalized leptonic matrix element as defined in~\cite{Beneke:2020vnb}.
The factorization for $B\to D^*$ is obtained by replacing $E_D\to E_{D^*} \; \epsilon^*\cdot v$. In full generality, the hard-scattering kernels and form factor $\zeta^{BD^{(*)}}$ depend on the lepton energy $E_\ell$ and $q^2 = (p_\ell + p_{\nu_\ell})^2$ and we defined as $H_{\rm sl}^{(*)}\equiv H_{\rm sl,-} \pm H_{\rm sl,+}$, where the upper (lower) sign applies to $D\;(D^*)$. Comparing the semi-leptonic factorization in \eqref{eq:QEDFsemi} with \eqref{eq:factscet1}, we observe that they are analogous under the replacements $E_L f_L \Phi_L \to E_{\rm sl} Z_\ell$ and $H \to H_{\rm sl}$, with the same HQET form factor $\zeta^{BD}$, in the appropriate kinematic limits. In addition, the hard Wilson coefficient that appears in the full amplitude is replaced by $C_2 \to C_{\rm sl}$.

\subsubsection{Physical form factor}
Similar to~\cite{Beneke:2020vnb}, for charged $L$ we replace the HQET form factor with the corresponding physical form factor $\mathcal{F}^{BD^{(*)}}$ obtained from the non-radiative semi-leptonic amplitude at the kinematic point $q^2 = m_L^2$ and at the maximal lepton energy at this point: $E_\ell^{\rm max} = (E_L + \sqrt{E_L^2 - m_L^2})/2$. We define
\begin{align}
\label{eq:curlyFexp}
\mathcal{F}^{BD}  \equiv \lim_{E_\ell \to E_\ell^{\rm max}}\frac{\mathcal{A}^{{\rm sl},D}_{\text{non-rad}}}{G_F/\sqrt{2} V_{cb} 4 i E_{D} E_{\rm sl}} \ , 
\end{align}
and equivalently for $D^*$ by replacing $E_D\to E_{D^*} \epsilon^*\cdot v$. For $q^2=m_L^2$, the spinor product for massless leptons becomes
\begin{equation}
    E_{\rm sl}(q^2=m_L^2) =  \sqrt{E_\ell^{\rm max}-E_\ell} \, \frac{8 E_\ell^{\rm max} \sqrt{E_\ell^{\rm max}(4E_\ell E_\ell^{\rm max} - m_L^2)}}{4(E_\ell^{\rm max})^2 - m_L^2} \ ,
\end{equation}
which goes to zero in the strict limit $E_\ell \to E_\ell^{\rm max}$, as the non-radiative amplitude goes to zero at the kinematic endpoint. The physical form factor $\mathcal{F}^{BD^{(*)}}$ however remains finite in this limit. Comparing with \eqref{eq:QEDFsemi}, we find
\begin{equation}
\label{eq:curlyF}
\mathcal{F}^{BD^{(*)}} = C_{\rm sl}\, Z_{\ell} \,\zeta^{BD^{(*)}}(E_\ell,q^2) H^{(*)}_{\rm sl}(E_\ell,q^2,z) \ .
\end{equation}
Finally, the relevant semi-leptonic hard-scattering kernels are
\begin{align}
H_{\rm sl,-}^{(1)} = &- Q_d Q_\ell \biggl\{\frac{L_b^2}{2} + L_b \biggl(1-2\ln(1-z) \biggr) + h(1-z) + \frac{\pi^2}{12} + 5 \biggr\} \nonumber \\
&+ Q_u Q_\ell \biggl\{\frac{L_c^2}{2}-3L_\nu +3L_b -2L_c \biggl(1-i\pi +\ln \Bigl(\frac{1-z}{z} \Bigr) \biggr) + g\Bigl(1-\frac{1}{z}\Bigr) + \frac{\pi^2}{12}-6 \biggr\} \nonumber \\
& + Q_u Q_d f(z) -Q_d^2 \biggl(\frac{3}{2}L_b +2 \biggr) -Q_u^2 \biggl(\frac{3}{2}L_c +2 \biggr) \ ,\\
H_{\rm sl,+}^{(1)} =& -Q_d Q_\ell \sqrt{z} \;w\big(1-z\big) - Q_u Q_d\sqrt{z} \frac{\ln z}{1-z} \ ,
\end{align}
where we used $E_\ell = m_b (1-z)/2$ for consistency as in the short-distance kernels we use the quark masses. 

Replacing now the HQET form factor $\zeta^{BD^{(*)}}$ using \eqref{eq:curlyF}, we find the factorization formula: 
\begin{eqnarray}\label{eq:qedffinal}
C_2 \left\langle D^{+} L^- | Q_2 |\bar B\right\rangle &=& 4i E_D E_L \frac{C_2}{C_{\rm sl}}\,
\mathcal{F}^{BD}(m_L^2) \, \int_0^1 du \, 
T(u,z) f_{L} \frac{\Phi_{L}(u)}{Z_\ell} 
 \, , 
\end{eqnarray}
where we defined
\begin{equation}
\label{eq:Tredefinition}
T^{(*)}(u)\equiv \frac{H^{(*)}(u)}{H^{(*)}_\textup{sl}} \ .
\end{equation}
The factorization formula for $\bar{B}\to D^{+*}L^-$ is obtained by replacing $E_D\to E_{D^*} \; \epsilon^*\cdot v$. In \eqref{eq:qedffinal}, we multiplied with the hard Wilson coefficient $C_2$ to make the different contributions explicit. We note that the drawback of \eqref{eq:qedffinal} is that the physical form factor $\mathcal{F}^{BD}$ includes electroweak-scale physics. On the other hand, the benefit of introducing this physical form factor is that \eqref{eq:curlyFexp} provides an explicit prescription allowing to extract it from experimental data. The procedure applied here is similar to the standard procedure in QCD alone, where the HQET $B\to D^{(*)}$ form factor is replaced by the full QCD form factor. This can be seen explicitly, when taking the limit $\alem\to 0$ and redefining the form factors in terms of the kinematical factors: 
\begin{align}
\hat{\mathcal{F}}^{BD}& \equiv \frac{4 E_D E_L}{m_B^2 - m_D^2} \mathcal{F}^{BD}  \underset{\alem\to\; 0}{\to}  F_0^{BD}
\ ,\label{eq:hatcurlf} \\
\hat{\mathcal{F}}^{BD^*} &\equiv -\frac{2 E_{D^*} E_L}{m_{D^*} m_B} \mathcal{F}^{BD^*} \underset{\alem\to\; 0}{\to}   A_0^{BD^{*}} 
\label{eq:hatcurlfstar} \ , 
\end{align}
where $F_0$ and $A_0$ are the QCD form factors defined in \cite{Beneke:2000ry}. 
The one-loop electromagnetic correction to the hard-scattering kernel is given by
\begin{align}
\label{eq:T2}
T^{(1)} = &\phantom{+} Q_d^2\biggl\{ 2L_b \ln u - h\bigl(u(1-z)\bigr) + h\bigl(1-z\bigr) \biggr\} \nonumber \\
&+ Q_u^2\biggl\{ -3L_\nu +3L_b - L_c(3-2\ln \bar{u}) -h\Bigl(\bar{u}\Bigl(1-\frac{1}{z}\Bigr)\Bigr) +g\Bigl(1-\frac{1}{z}\Bigr) -11 \biggr\} \nonumber \\
& +  Q_d Q_u\biggl\{ -3L_\nu -2L_b \ln \bar{u} -2L_c \ln u +g\bigl(\bar{u}(1-z)\bigr) + g\Bigl(u\Bigl(1-\frac{1}{z}\Bigr)\Bigr) \nonumber \\
&-h\bigl(1-z\bigr) - g\Bigl(1-\frac{1}{z}\Bigr)-11 \biggr\} \nonumber \\
&-\sqrt{z}\biggl[ Q_d^2 \Bigl(w\bigl(u(1-z)\bigr) - w(1-z)\Bigr) + Q_u^2 \frac{1}{z}w\Bigl(\bar{u}\Bigl(1-\frac{1}{z}\Bigr)\Bigr) + Q_d Q_u w(1-z) \, \biggr] \ ,
\end{align}
and the corresponding $T^{*}$ is obtained by setting $\sqrt{z}\to -\sqrt{z}$. Unlike $H_-$, the hard-scattering kernel $T$ in~\eqref{eq:T2} is free from collinear divergences when taking the limit $z\to 0$. In this limit,  \eqref{eq:T2} reduces to (73) in \cite{Beneke:2020vnb}.


\section{QED corrections to branching fractions and ratios}
\label{sec:pheno}

\begin{table}[t]
	\begin{center}
		\begin{tabularx}{0.98\textwidth}{|c C C C|}
	\hline 
	\multicolumn{4}{|c|}{Coupling constants and masses [GeV]}\\
		\hline 
  $\alem(m_Z)=1/127.96$  & $\alpha_s(m_Z) = 0.1181$ & $m_B = 5.279$ & $m_Z=91.1876$ 
  \\ \hline
		\end{tabularx}
				\vskip 1pt
\begin{tabularx}{0.98\textwidth}{|C C C C|}
\hline
	\multicolumn{4}{|c|}{Decay constants [MeV] and masses}\\
		\hline 
$f_\pi= 130.2 \pm 1.4$  & $f_K=155.6 \pm 0.4$  & $m_b = 4.78$ GeV & $m_c = 1.67$ GeV 
\\  	\hline
\end{tabularx}
\vskip 1pt
\begin{tabularx}{0.98\textwidth}{|C C|}
\hline
\multicolumn{2}{|c|}{CKM parameters}  \\ \hline
$|V_{ud}|= 0.97370 \pm 0.00014$  &  $|V_{us}|=0.2245 \pm 0.0008$\\ 
\hline
\end{tabularx}
\vskip 1pt
	\begin{tabularx}{0.98\textwidth}{|C C C C|}
	\hline
	\multicolumn{4}{|c|}{Wilson coefficients and coupling constants at $\nu=4.8$ GeV}\\
		\hline 
  $C_1^{\rm QCD}=-0.258$ & $C_2^{\rm QCD}=1.008$    & $\alem=1/132.24$  & $\alpha_s^{(4)} = 0.21617  $ 
  \\ \hline\end{tabularx}
		\vskip 1pt
		\begin{tabularx}{0.98\textwidth}{|C C C|}
\hline	\multicolumn{3}{|c|}{Parameters of distributions amplitudes at $\mu=1$ GeV}\\
		\hline 	
$a_2^{\pi}=0.15$ & $a_1^{\bar{K}}=0.06$ & $a_2^{\bar{K}} = 0.14$ \\ \hline
$a_2^{\rho}=0.17$ & $a_1^{\bar{K^*}}=0.06$ & $a_2^{\bar{K^*}} = 0.16$ \\ \hline
\end{tabularx}
		\vskip 1pt
		\begin{tabularx}{0.98\textwidth}{|C |}
\hline	\multicolumn{1}{|c|}{Coupling constants at $\mu=1$ GeV}\\
		\hline 	
 $\alem=1/134.05$
 \\ \hline
\end{tabularx}
\caption{Inputs used to obtain QCD NNLO predictions and QED effects. The Gegenbauer coefficients of the $\pi$ and $K$ are taken from \cite{Bali:2019dqc} and evolved to $1\,$GeV with LL accuracy. The $\rho$ and $K^*$ Gegenbauer coefficients are taken from \cite{Straub:2015ica}, which uses Lattice QCD results from \cite{Arthur:2010xf}.
We use $\alpha_s^{(4)}$ with four flavours and the three-loop running, while the Wilson coefficients were matched to the full SM at the electroweak scale $\mu_0= m_W$. The quark masses are to be understood as two-loop pole masses.}
\label{tab:inputs}
	\end{center}
\end{table}
%
%
In this section, we discuss the phenomenological implications of the QED effects on $B \to D^{(*)} L$ decays, where we focus on $L = \pi, K$. Our formalism also holds for vector meson final states $L=\rho, K^*$ and baryon decays, which we briefly discuss at the end of Sec.~\ref{sec:qedampl}. We divide the QED effects in three groups arising at different scales:
\begin{itemize}
    \item corrections to Wilson coefficients above the scale $m_b$,
    \item corrections to the kernels, form factors and LCDA from scales between $m_b$ and $\Lambda_{\rm QCD}$,
    \item corrections to the decay rate from ultrasoft photon radiation below $\Lambda_{\rm QCD}$.
\end{itemize}
The first two corrections contribute at the amplitude level and can be included via 
\begin{equation}\label{eq:fact}
    \mathcal{A}(B \to D^{(*)} L) = \mathcal{A}_{BD^{(*)}} \;a_1(D^{(*)} L) \ ,
\end{equation}
 where $a_1$ parametrizes the colour-allowed tree-amplitude and
\begin{equation}\label{eq:AtoDLfirst}
    \mathcal{A}_{BD} =  i  \frac{G_F}{\sqrt{2}}V^*_{uD} V_{cb} f_L  \; 4E_L E_D\mathcal{F}^{BD}(m_L^2) \ .
\end{equation}
A similar expression holds for $D^*$ with the replacement $E_D \to E_{D^*} \; \eps^* \cdot v$. In QCD, taking the limit $\alem \to 0$, $\mathcal{A}_{BD}$ in~\eqref{eq:AtoDLfirst} reduces to 
\begin{align}
 A_{BD}&\equiv i \frac{G_F}{\sqrt{2}}V^*_{uD} V_{cb} f_L (m_B^2 - m_D^2)F_0^{BD}(m_L^2) \ , \end{align}
while for $D^*$, we have 
\begin{align}
 A_{BD^{*}}&\equiv -i \frac{G_F}{\sqrt{2}}V^*_{uD} V_{cb} f_L  2m_{D^*}\; \epsilon^* \cdot p_B A_0^{BD^*}(m_L^2) \,.
\end{align}
To parametrize the QED effects, we can now factor out this QCD expression by writing
\begin{equation}
\label{eq:AtoDL}
    \mathcal{A}(B \to D^{(*)} L) =  A_{BD^{(*)}} \, {\biggl(\frac{\hat{\mathcal{F}}^{BD^{(*)}}}{F^{BD^{(*)}}_0}\biggr)} \, \left[a_1^{\rm QCD}(D^{(*)} L) + \delta a_1(D^{(*)} L)\right] \ ,
\end{equation}
in terms of the reduced form factors $\hat{\mathcal{F}}_{BD^{(*)}}$ defined in \eqref{eq:hatcurlf} and \eqref{eq:hatcurlfstar}. In addition, for $B\to D^*$, we have $F_0^{BD^*} \equiv A_0^{BD^*}$. The shift $\delta a_1$ encodes the $\mathcal{O}(\alem)$ QED correction coming from the Wilson coefficient, the hard-scattering kernel, and the LCDA, respectively:
\begin{align}
\label{eq:deltaa1}
   \delta a_1(D^{(*)} L) &\equiv \delta a_1^{\rm WC}(D^{(*)} L)+ \delta a_1^{\rm K}(D^{(*)} L) + \delta a_1^{\rm L}(D^{(*)} L) \,.
\end{align}
Due to the replacement of the HQET form factor by $\mathcal{F}_{BD^{(*)}}$, this includes QED effects from the semi-leptonic Wilson coefficient $C_{\rm sl}$ in $\delta a_1^{\rm WC}$, from $H_{\rm sl}$ in $\delta a_1^K$ and the leptonic matrix element $Z_\ell$ in $\delta a_1^L$, as can be seen from \eqref{eq:qedffinal}.
However, $\delta a_1$ does not include QED corrections to the form factor, which are contained in the ratio  $\hat{\mathcal{F}}_{BD^{(*)}}/F_0^{BD^{(*)}} = 1 + \mathcal{O}(\alem)$.\footnote{Explicitly, the QED corrections to the physical form factor defined in \eqref{eq:curlyF} are
\begin{align*}
 \hat{\mathcal{F}}^{BD^{(*)}}(m_L^2) & = F_0^{BD^{(*)}}(m_L^2) \;\left(1+ \delta a^{\rm WC}_{\rm sl} + \delta a^{\rm K}_{\rm sl} + \delta a^{\rm L}_{\rm sl} + \delta a^{\rm F}_{\rm sl}  \right)  \ ,
\end{align*}
which includes the QED corrections to the Wilson coefficient $C_{\rm sl}, H_{\rm sl}, Z_\ell$ and HQET form factor $\zeta^{BD^{(*)}}$, respectively. 
}
These QED corrections cancel in ratios with semi-leptonic decays, for which we provide numerical estimates in Sec.~\ref{subsec:slratios}.
For our numerical analysis, we use the inputs given in Table~\ref{tab:inputs}. For completeness, we also evaluate the pure QCD NNLO $a_1(D L)$ using \cite{Huber:2016xod} with our inputs: 
\begin{equation}
\begin{split}
\label{eq:a1qcd}
a_1^{\rm QCD}(D^+K^-)&= 1.008 + [0.024 + 0.009 i]_{\rm NLO} + [0.029 + 0.029 i]_{\rm NNLO} \\
&= 1.061^{+0.017}_{-0.016}+ 0.038^{+0.025}_{-0.014} i\,,
\end{split}
\end{equation}
where the uncertainty is fully dominated by the scale variation $m_b/2<\mu <2m_b$ around the central value $\mu=m_b$. Here we used the Gegenbauer coefficients from \cite{Bali:2019dqc} evolved to $m_b$ with NLL accuracy. The uncertainty coming from the charm mass is negligible. Even when allowing very conservative variations of both $1.3\; {\rm GeV}
<m_c <1.7$ GeV and $4.5 \; {\rm GeV}<m_b<4.9$ GeV, it is one order of magnitude smaller than the one from the scale variation in \eqref{eq:a1qcd}.
Our result differs slightly from \cite{Huber:2016xod} (see also \cite{Bordone:2020gao}) due to different inputs and because we do not re-expand the Wilson coefficients as a series in $\alpha_s$, but treat the values provided in Table~\ref{tab:inputs} as inputs. 
Numerical values of $a_1$ for final states other than $D^+K^-$ are summarized in Table~\ref{tab:a1vals}. We note that the numerical differences between $D$ and $D^*$ final states are small.

\subsection{QED corrections to $a_1$}
\label{sec:qedampl}

At the amplitude level, at linear order in $\alem$, there are four separate QED corrections, three of which contribute to $\delta a_1$ defined in \eqref{eq:deltaa1}. We remind the reader that by definition the QED corrections in $\mathcal{F}^{BD^{(*)}}$ are not included in the shift $\delta a_1$. The QED correction to the Wilson coefficient was obtained in \cite{Beneke:2020vnb} (see \cite{Huber:2005ig,Bobeth:2003at}). As here we only deal with charged $L$ mesons, for which only $C_2$ contributes, we need
 \begin{equation}
     \delta a_1^{\rm WC}(D^{(*)}L) = \delta C_2 - \delta C_{\rm sl} C_2^{\rm tree} =  -0.39 \cdot 10^{-2}\,,
 \end{equation}
where $\delta C_{\rm sl}$ is the contribution from the semi-leptonic Wilson coefficient (as seen from ~\eqref{eq:qedffinal} and discussed below~\eqref{eq:deltaa1}). Neglecting $\mathcal{O}{(\alpha_s \alem)}$ corrections for consistency, we have defined the tree-level Wilson coefficient $C_2^{\rm tree} = 1.03$ (see~(104) in~\cite{Beneke:2020vnb}). The kernel contribution $\delta a_1^{\rm K}$ is
\begin{equation}
    \delta a_1^{\rm K}(D^{(*)+}L^-) \equiv \frac{\alem(\mu)}{4\pi} C_2^{\rm QCD}(\nu) \int_0^1 du \; T^{(*)(1)}(u, z)\phi_L(u) \ ,
\end{equation}
where $\phi_L$ is the QCD LCDA for which we use the standard Gegenbauer expansion. $\mathcal{O}(\alem)$ QED effects from the QED-generalized LCDA are parametrized by $\delta a_1^{\rm L}$:
\begin{equation}
    \delta a_1^{\rm L} \equiv C_2^{\rm QCD}(\nu) \int_0^1 du \left(\frac{\Phi_L(u)}{Z_\ell}- \phi_L(u)\right) \ .
    \end{equation}
The QED corrections from the QED generalized light-meson LCDA $\delta a_1^{\rm L}$ require long-distance calculations. The generalized QED light-meson LCDA $\Phi_L$ is discussed in detail in \cite{BLCDApaper}. In principle, it has to be obtained either by lattice QCD simulations or directly from experiment. Although some progress has been made including QED corrections to the form factors of light mesons (see e.g. \cite{Carrasco:2015xwa}), such calculations are not available yet for $B\to D^{(*)}$ form factors, nor for the $\pi$ and $K$ LCDAs. In the following, to discuss the QED effects on $a_1$, we therefore neglect the corrections from $\delta a_1^{\rm L}$.

For the convoluted kernel results, we truncate the Gegenbauer expansion at the second moment, and write
\begin{align}
\label{eq:Vdef}
 \int_0^1 \!\! du \; T^{(*)(1)}(u,z) \phi_L(u) &=  V_{0}^{(*)(1)}(\mu,z) + a_1^L(\mu) \, V_{1}^{(*)(1)}(\mu,z) +  a_2^L(\mu) \, V_{2}^{(*)(1)}(\mu,z) \, , 
\end{align}
where $a_{1,2}^L$ are the first two Gegenbauer coefficients of the QCD light meson LCDA. Analytic expressions for $V_{i}^{(*)(1)}(\mu, z)$ are given in Appendix~\ref{sec:app:expr}.
At $\mu=1$ GeV and $\nu=4.8$ GeV, we find 
\begin{align}\label{eq:a1withgeg}
    \delta a_1^{\rm K}(D^{+} L^- ) &=\frac{\alem(\mu)}{4\pi} \Bigl[ -1.81 - 8.71 i -  (0.49 + 1.18 i)  a_1^L(\mu)-(0.92 + 1.97 i) a_2^L(\mu) \Bigr] \ ,\nonumber \\
    \delta a_1^{\rm K}(D^{+ *} L^- ) &=\frac{\alem(\mu)}{4\pi} \Bigl[ -1.69 - 7.30 i + (0.07 - 0.76 i) a_1^L(\mu)  -  (0.51 + 1.96 i) a_2^L(\mu) \Bigr] \ ,
\end{align}
where we kept the Gegenbauer coefficients unevaluated. Using their numerical values at $\mu=1$ GeV from Table~\ref{tab:inputs}, we obtain 
\begin{align}
    \delta a_1^{\rm K}(D^+ \pi^- ) &= \frac{\alem(\mu)}{4\pi}\biggl[-1.95-9.01i \biggr] = (-0.12-0.53i)\cdot 10^{-2}\,, \\
    \delta a_1^{\rm K}(D^+ K^- ) &= \frac{\alem(\mu)}{4\pi}\biggl[-1.97-9.06i \biggr] = (-0.12-0.54i)\cdot 10^{-2}\,, \\
    \delta a_1^{\rm K}(D^{*+} \pi^- ) &= \frac{\alem(\mu)}{4\pi}\biggl[-1.77-7.60i \biggr] = (-0.10-0.45i)\cdot 10^{-2}\,, \\
    \delta a_1^{\rm K}(D^{*+} K^- ) &= \frac{\alem(\mu)}{4\pi}\biggl[-1.76-7.62i \biggr] = (-0.10-0.45i)\cdot 10^{-2}\,.
\end{align}
Comparing to QCD, we find that the imaginary part is about $15\%$ of the QCD imaginary part in \eqref{eq:a1qcd}, while the real part is around four times smaller.

In Table~\ref{tab:a1vals}, we summarize our results adding $\delta a_1^{\rm K} $ and $\delta a_1^{\rm WC}$ to the NNLO QCD expressions, the latter obtained using the ancillary files of \cite{Huber:2016xod} and the inputs in Table~\ref{tab:inputs}. Comparing to QCD, we find that the QED effects are somewhat smaller than the QCD uncertainty. Therefore, the quoted uncertainty in the final column is that of the QCD NNLO prediction and we do not assign an additional QED uncertainty. The QED effects on the real part and absolute values of $a_1$ are dominated by $\delta a_1^{\rm WC}$. Our results also apply to non-leptonic two-body $B_s \to D_s$ and $\Lambda_b\to \Lambda_c$ decays with
\begin{equation}
    a_1(D^+L^-) = a_1(D_s^+ L^-) = a_1(\Lambda_c^+ L^-) \ ,
\end{equation}
which holds in QCD and also for the QED correction $\delta a_1$. Finally, we stress that \eqref{eq:a1withgeg} holds also for light vector-meson final states with $L=\rho, K^*$. Taking the Gegenbauer moment coefficients as listed in Table~\ref{tab:inputs}, we find
\begin{align}
    \delta a_1^{\rm K}(D^+ \rho^- ) &= (-0.12-0.54i)\cdot 10^{-2}\,, \\
    \delta a_1^{\rm K}(D^+ K^{*-} ) &=  (-0.12-0.54i)\cdot 10^{-2}\, . 
\end{align}
However, as finite-width effects may play an important role (see \cite{Huber:2020pqb}), we do not discuss these results further in detail.  

\begin{table}[t]
\begin{center}
\begin{tabular}{lccccc}
\hline \hline
&&&&\\[-0.5cm]
$|a_1(D^{(\ast)+}L^-)|$ & ${\rm LO}$ & ${\rm NLO}$ & ${\rm NNLO}$ & +QED NLO \\
\hline
&&&&\\[-0.3cm]
  $|a_1(D^{+}\pi^-)|$
& $\phantom{-}1.008$
& $\phantom{-}1.032^{+0.024}_{-0.018}$
& $\phantom{-}1.064^{+0.019}_{-0.017}$
& $\phantom{-}1.059^{+0.019}_{-0.017}$ \\ \addlinespace
  $|a_1(D^{\ast+}\pi^-)|$
& $\phantom{-}1.008$
& $\phantom{-}1.031^{+0.023}_{-0.018}$
& $\phantom{-}1.063^{+0.020}_{-0.017}$
& $\phantom{-}1.058^{+0.020}_{-0.017}$ \\ \addlinespace
\hline
&&&&\\[-0.3cm]
  $|a_1(D^{+}K^-)|$
& $\phantom{-}1.008$
& $\phantom{-}1.032^{+0.024}_{-0.018}$
& $\phantom{-}1.062^{+0.017}_{-0.016}$
& $\phantom{-}1.057^{+0.017}_{-0.016}$ \\ \addlinespace
  $|a_1(D^{\ast+}K^-)|$
& $\phantom{-}1.008$
& $\phantom{-}1.031^{+0.023}_{-0.018}$
& $\phantom{-}1.061^{+0.017}_{-0.016}$
& $\phantom{-}1.056^{+0.017}_{-0.016}$ \\ \addlinespace
\hline \hline

\end{tabular}
\end{center}
\caption{Results for $|a_1(D L)|$ at LO, NLO and NNLO using \cite{Huber:2016xod} for the NNLO expressions with inputs from Table~\ref{tab:inputs}. In the last column, we added the QED NLO corrections $\delta a_1^{\rm K}$ and $\delta a_1^{\rm WC}$. }
\label{tab:a1vals}
\end{table}

\subsection{Ultrasoft effects}

In the previous section, we considered the QED effects to the non-radiative amplitude. At the level of the decay rate also ultrasoft-photon effects have to be included to render the observable IR finite. We define the ultrasoft-photon-inclusive rate $\Gamma[\bar{B}\rightarrow D^{(*)} L](\Delta E)$ as the $\Gamma[\bar{B}\to D^{(*)} L + X_s]$ where $X_s$ consists of photons and electron-positron pairs with total energy $\Delta E \ll \Lambda_{\rm QCD}$. For general $D^{(*)}L$ final states, the ultrasoft-photon-inclusive rate factorizes as
\begin{equation}
\Gamma[\bar{B}\rightarrow D^{(*)} L](\Delta E) = |\mathcal{A}(\bar{B}\rightarrow D^{(*)} L) |^2 \; \mathcal{S}_{\otimes}(\Delta E)  \ ,
\end{equation}
where $\otimes = (Q_D, Q_L)$ and
\begin{equation}
\label{eq:Smu}
S_\otimes(\Delta E) = 1 + \frac{\alpha_\textup{em}}{4\pi}\biggl[8 b_\otimes \ln \biggl(\frac{\mu}{2 \Delta E} \biggr) + 4 F_\otimes \biggr] + \mathcal{O}\left(\alem^2\right)\ .
\end{equation}
In the double-logarithmic approximation $S_\otimes$ exponentiates to 
\begin{equation}
    S_\otimes = {\rm exp} \left(\frac{\alem}{4\pi} S_\otimes^{(1)}\right) \ ,
\end{equation}
which will be used below. 
For $\bar{B}\to D^+ L^-$, the functions $b_{(+,-)}$ and $F_{(+,-)}$ are given by \cite{Baracchini:2005wp}  
\begin{align}
b_{(+,-)} &= 1 - \frac{4-\Delta_+^2 - \Delta_-^2 + 2\beta^2}{8 \beta}\biggl[\ln \Bigl(\frac{\Delta_+ + \beta}{\Delta_+ - \beta} \Bigr) + \ln \Bigl(\frac{\Delta_- + \beta}{\Delta_- - \beta} \Bigr) \biggr] \label{eq:botimes} \\
F_{(+,-)} &= \frac{\Delta_+}{2 \beta}\ln \Bigl(\frac{\Delta_+ + \beta}{\Delta_+ - \beta} \Bigr) + \frac{\Delta_-}{2 \beta}\ln \Bigl(\frac{\Delta_- + \beta}{\Delta_- - \beta}  \Bigr) + \frac{4-\Delta_-^2 - \Delta_+^2 + 2\beta^2}{4 \beta} \nonumber \\
 &\times \biggl[\rm{Re}\Bigl\{\textup{Li}_2\Bigl(-\frac{\beta}{\Delta_-} \Bigr) -\textup{Li}_2\Bigl(\frac{\beta}{\Delta_-} \Bigr)+\frac{1}{2}\textup{Li}_2\Bigl(\frac{\Delta_- + \beta}{2\Delta_-} \Bigr)-\frac{1}{2}\textup{Li}_2\Bigl(\frac{\Delta_--\beta}{2\Delta_-} \Bigr) \Bigr\} \nonumber\\
&- \frac{\ln 2}{2}\ln \Bigl(\frac{\Delta_- + \beta}{\Delta_- - \beta} \Bigr)+\frac{1}{4}\ln^2 \Bigl(1+\frac{\beta}{\Delta_-} \Bigr)-\frac{1}{4}\ln^2 \Bigl(1-\frac{\beta}{\Delta_-} \Bigr) + (\Delta_- \rightarrow \Delta_+)    \biggr], \label{eq:Fotimes}
\end{align}
where  $\beta\equiv \sqrt{(1-(r_L+r_D)^2)(1-(r_L-r_D)^2)}$, 
and we defined $r_L \equiv m_L/m_B$, $r_D\equiv m_D/m_B$, and $\Delta_\pm \equiv 1 \pm r_L^2 \mp r_D^2$. Expression \eqref{eq:Smu} agrees with the energy-dependent correction factor (5) in \cite{Baracchini:2005wp}, when setting $\mu=m_B$ and dropping the virtual contributions $H_{12}$ and $N_{12}$ in that reference. The conceptual differences between the approach of \cite{Baracchini:2005wp} and ours are discussed in detail in Sec. 6 of \cite{Beneke:2020vnb}.

The scale $\mu$ in \eqref{eq:Smu} is related to the IR divergence of the non-radiative amplitude \cite{Beneke:2020vnb}.
In the present case, the generalized form factor and LCDA must be understood as properly IR subtracted and thus as scheme-dependent quantities. As we can compute only the UV scale dependence of these objects with perturbative methods, we are forced to choose $\mu$ in \eqref{eq:Smu} to be of the order of the collinear scale to correctly resum all large logarithms. In the following we restrict ourselves to the resummation of double logarithms only, except for logarithms in $\Delta E$, as was done in~\cite{Beneke:2020vnb}. We note that these double logarithmic contributions arise entirely from QED effects. After soft rearrangement, all large double logarithms are removed from the heavy-to-heavy current. This can be seen by looking at the UV poles in~\eqref{eq:Zhh}: the potentially large logarithm in the second line involves the off-shell regulator, and is cancelled after dividing by $R_{\bar{c}}^{(Q_L)}$. Hence, we only have to consider the two Sudakov factors of the hard-scattering kernel ($S_H$), and the QED generalized light-meson LCDA ($S_{\bar{C}}$, see Eq. (4.26) in~\cite{Beneke:2019slt}):
\begin{align}
    S_H(\mu,\mu_b) &= \exp \left\{-\frac{\alem}{2\pi} Q_L^2 \ln^2 \frac{\mu}{\mu_b}\right\}, \\
    S_{\bar{C}}(\mu,\mu_c;E_L)
    &= \exp \left\{-\frac{\alem}{2\pi} Q_L^2 \left( \ln^2 \frac{\mu_c}{2E_L} - \ln^2\frac{\mu}{2E_L} \right) \right\} \,.
\end{align}
Multiplying the two gives the universal Sudakov factor that resums large double-logarithms between the hard and the collinear scale:
\begin{equation}
\label{eq:sudakcompact}
S_{L}(\mu_b,\mu_c) = \exp\left\{-\frac{\alpha_\textup{em}}{2\pi}Q_{L}^2 \ln^2 \frac{\mu_c}{\mu_b}\right\} \,,
\end{equation}
where we used 
\begin{align}
    \ln^2\frac{\mu}{\mu_b} + \ln^2 \frac{\mu_c}{2E_L} - \ln^2\frac{\mu}{2E_L}  = \ln^2 \frac{\mu_b}{\mu_c} -2 \ln\frac{\mu}{\mu_c}\ln\frac{\mu_b}{2E_L} \,,
\end{align}
and dropped the second term on the right-hand side in the double-logarithmic approximation, since $2E_L =\mathcal{O}(\mu_b)$.

Combining this with the exponentiated ultrasoft function at $\mu=\mu_c$, allows us to write the ultrasoft-photon-inclusive width as
\begin{equation}
\label{eq:softphotonincl}
    \Gamma\left[\bar{B}\to D^{(*)}L\right](\Delta E) = \Gamma^{(0)}\left[\bar{B}\to D^{(*)}L\right] U(D^{(*)}L)\ ,
\end{equation}
where the non-radiative amplitude $\Gamma^{(0)}$ is defined as the square of the factorized virtual $\bar{B} \to D^+ L^-$ amplitude, $|\mathcal{A}(\bar{B}^0\to D^{+(*)}L^-)|^2$, divided by the factor in \eqref{eq:sudakcompact}. The $U$ factor is given by
\begin{equation}
\label{eq:Usoft}
U(D^{(*)} L) \equiv |e^{S_{L}(\mu_b,\mu_c)}|^2 e^{\frac{\alpha_\textup{em}}{4\pi}S^{(1)}_{\otimes}}= \biggl(\frac{2\Delta E}{m_B}\biggr)^{-\frac{\alpha_\textup{em}}{\pi}2b_\otimes} \times t_\otimes(\mu_c) \ , 
\end{equation}
with
\begin{equation}
\label{eq:totimes}
    t_\otimes(\mu_c)= {\rm exp}\left\{ \frac{\alpha_{\rm em}}{\pi} \left[2b_\otimes \ln\frac{\mu_c}{m_B} - Q_{L}^2 \ln^2\frac{\mu_c}{m_B} + F_\otimes\right]\right\} \ .
\end{equation}
As discussed, we restrict ourselves to the double-logarithmic approximation (except for logarithms in $\Delta E$). Expanding up to leading order in $m_L/m_B = \mathcal{O}(\Lambda_{\rm QCD}/m_B)$, gives
\begin{align}
\label{eq:botimesexp}
b_{(+,-)} &= 1+\ln \biggl(\frac{m_D m_L}{m_B^2-m_D^2} \biggr) \ ,
\end{align}
and
\begin{equation}
    F_{(+,-)} = -\ln^2 \frac{m_L m_B}{m_B^2-m_D^2}
    \ ,
\end{equation}
where in the expression for $F_{(+,-)}$ we only kept the double-logarithmic terms. In this approximation \eqref{eq:totimes} reduces to
\begin{equation}
    t_{(+,-)}(\mu_c)=  {\rm exp}\left\{ \frac{\alpha_{\rm em}}{\pi} \left[-\ln^2\left(\frac{m_L}{\mu_c}\frac{m_B^2}{m_B^2-m_D^2}\right)  \right] 
    \right\}\ .
\end{equation}
Since $m_L = \mathcal{O}(\mu_c)$, no large double logarithms remain, $t_\otimes=1$ in this approximation, and as expected, the $\mu_c$ dependence in \eqref{eq:Usoft} cancels. 

Numerically, some of the sub-leading logarithmic terms in $t_{\otimes}$ turn out to be at the percent level when setting $\mu_c=1$ GeV. A consistent treatment of sub-leading logarithms would however require a non-perturbative matching of SCET onto an effective theory of point-like mesons, which is beyond the scope of the present work. In the remainder, we therefore consider ultrasoft effects following from $U(D^{(*)}L)$ as defined in \eqref{eq:Usoft} with $t_{\otimes}=1$ and $b_\otimes$ in \eqref{eq:botimesexp}, similar as in the QCD$\times$QED factorization of the $B$ to two light meson decay \cite{Beneke:2020vnb}. However, as these ultrasoft contributions are the dominant QED effects, we take the difference between the double-logarithmic approximation of \eqref{eq:Usoft} with $t_\otimes=1$ and the full $t_\otimes(\mu_c)$ using \eqref{eq:botimes} and \eqref{eq:Fotimes} evaluated at $\mu_c=1$ GeV as a two-sided conservative uncertainty on $U(D^{(*)}L)$. Taking $\Delta E = 60\,$MeV as in \cite{Beneke:2020vnb, Beneke:2019slt}, we find 
\begin{align}
\label{eq:usoftval}
U(D^+ K^-) & = 0.960 \pm 0.001 \ ,  \\
U(D^+ \pi^-) &= 0.938 \pm 0.005 \ , \\
U(D^{*+} K^-) & = 0.962 \pm 0.001 \ ,  \\
U(D^{*+} \pi^-) &= 0.940 \pm 0.005 \ .\label{eq:usoftvallast}
\end{align}
Finally, we stress that in principle the semi-leptonic rate used to normalize our expressions also gets an ultrasoft correction $U(D^{(*)} \ell)$. We discuss this in more detail in the next section and note that we refer to contributions coming from $U$ as ultrasoft effects.

\subsection{Ratios of semi-leptonic and non-leptonic rates}
\label{subsec:slratios}
Previously, we focused on the QED corrections to $a_1$, and ignored contributions from $\mathcal{F}^{BD^{(*)}}/F_0^{BD^{(*)}}$ as these are presently not known. It is therefore of interest to consider ratios of non-leptonic over semi-leptonic decays as is generally done to suppress the effect of the form factors. In this way, the coefficients $|a_1(D^{(\ast)+}L^-)|$ can be extracted directly from experimental data allowing for a clean test of QCD and QCD$\times$QED factorization \cite{Beneke:2000ry,Cai:2021mlt,Neubert:1997uc}. 

The dominant QED effects are from (ultra)soft radiation. In the following, we assume that the non-leptonic rate always includes soft photons. Following the definition in \eqref{eq:softphotonincl} this is denoted by $\Gamma(\bar{B}_{d}\to D^{(\ast)+}L^-)(\Delta E)$. Contrary to QCD, we now define two ratios of non-leptonic versus semi-leptonic decays depending on whether or not in the semi-leptonic rate soft-photon radiation is included, or whether the non-radiative rate is considered.

Assuming that the non-radiative semi-leptonic rate was extracted from data, we define 
\begin{align} \label{eq:nonlep2semilep}
R_L^{(0),(\ast)}(\Delta E) &\equiv \frac{\Gamma(\bar{B}_{d}\to D^{(\ast)+}L^-)(\Delta E)}{d\Gamma^{(0)}(\bar{B}_{d}\to D^{(\ast)+}\ell^-\bar{\nu}_{\ell})/dq^2\mid_{q^2=m_L^2}}\, \nonumber \\
&= 6\pi^2\,|V_{uD}|^2 U(D^{(*)}L)\,f_L^2\,\left|a_1^{\rm QCD}(D^{(*)} L) + \delta a_1(D^{(*)} L)\right|^2\, X_L^{(\ast)}\, \nonumber \\
&\equiv \, R_L^{(\ast)}|_{\rm QCD} \left(1 + \delta_{\rm QED}(D^{(*)}L) + \delta_{\rm U}^{(0)}(\Delta E)\right) \ ,
\end{align}
where
\begin{equation}
\delta_{\rm QED}\equiv \frac{2 {\rm Re} (\delta a_1^{\rm WC} + \delta a_1^{\rm K}+\delta a_1^{\rm L})a_1^{\rm tree}}{|a_1^{\rm QCD}|^2} \,.
\end{equation}
The ultrasoft effects are  $\delta^{(0)}_{\rm U}\equiv U(D^{(*)}L)-1$ which can be obtained from \eqref{eq:usoftval}$-$\eqref{eq:usoftvallast}.  
Second, we consider the ratio where both the semi-leptonic and the non-leptonic rates are the soft-photon inclusive: 
\begin{align} \label{eq:nonlep2semilepII}
R_L^{(\ast)}(\Delta E) &\equiv \frac{\Gamma(\bar{B}_{d}\to D^{(\ast)+}L^-)(\Delta E)}{d\Gamma(\bar{B}_{d}\to D^{(\ast)+}\ell^-\bar{\nu}_{\ell})(\Delta E)/dq^2\mid_{q^2=m_L^2}}\, ,
\end{align}
which can be expressed as \eqref{eq:nonlep2semilep} but now also the ultrasoft factor $U(D^{(*)}\ell)$ should be included by replacing 
$U(D^{(*)}L)\to U(D^{(*)}L) / U(D^{(*)} \ell)$ and $\delta_{\rm U}^{(0)} \to \delta_{\rm U} \equiv U(D^{(*)}L)/U(D^{(*)} \ell)-1$. The latter only differs from zero due to the small differences in mass between $L$ and $\ell= \mu$. For $\Delta E = 60$ MeV we find 
\begin{align}
\delta_{\rm U}(D^+ \pi^-) &= \delta_{\rm U}(D^{+*} \pi^-) =  0.005 \pm 0.009 \ , \\
    \delta_{\rm U}(D^+ K^-) &=     \delta_{\rm U}(D^{+*} K^-)=  0.028 \pm 0.008 \ . 
\end{align}
In the ratios \eqref{eq:nonlep2semilep} and \eqref{eq:nonlep2semilepII} the form factor $\mathcal{F}^{BD^{(*)}}$ cancels. Neglecting $\delta a_1^{\rm L}$ in the following, $\delta_{\rm QED}$ only contains QED corrections to the kernel and the Wilson coefficients. We find:   
\begin{align}
\label{eq:deltaQED}
    \delta_{\rm QED}(D \pi) &= -0.90 \cdot 10^{-2} \,, \\
    \delta_{\rm QED}(D K) &= -0.90\cdot 10^{-2} \,, \\
    \delta_{\rm QED}(D^* \pi)& = -0.88 \cdot 10^{-2}\,, \\
    \delta_{\rm QED}(D^* K) &= -0.88 \cdot 10^{-2}\,.
\label{eq:deltaQEDlast}
\end{align}

For the QCD contribution, we include the form factor effects to the ratio $R_L$ which enter via 
\begin{equation}
\begin{split}
\label{eq:XLFull}
    X_L &= \frac{(m_B^2-m_D^2)^2}{[m_B^2 -(m_D-m_L)^2][m_B^2 -(m_D+m_L)^2]}\left|\frac{F_0(m_L^2)}{F_1(m_L^2)}\right|^2 \,,\\
    X_L^* &= [m_B^2 -(m_{D^*}-m_L)^2][m_B^2 -(m_{D^*}+m_L)^2]\frac{|A_0(m_L^2)|^2}{m_L^2 \sum_{i=0,\pm}|\mathcal{H}_i(m_L^2)|^2} \,,
\end{split}
\end{equation}
where the form factors $F_0, F_1$ and $A_0$ are defined in \cite{Neubert:1997uc} and $\mathcal{H}_0(q^2)$ and $\mathcal{H}_\pm(q^2)$ are the helicity amplitudes defined in the Appendix of~\cite{Neubert:1997uc}. Using the large-recoil relations for the form factors (see  \cite{Beneke:2000wa}), and expanding $X_L$ in powers of $m_L^2/m_B^2$ gives~\cite{Neubert:1997uc}
\begin{equation}
\begin{split}
\label{eq:approx}
    X_L &= 1 + \frac{4 m_L^2 m_B m_D}{(m_B^2-m_D^2)^2} \,,\\
    X^*_L &= 1 + \frac{4 m_L^2 m_B m_{D^*}}{(m_B^2-m_{D^*}^2)^2} - \frac{4 m_L^2}{(m_B - m_{D^*})^2} \,.
\end{split}
\end{equation}
Numerically, using the approximation \eqref{eq:approx} gives
\begin{equation}
\label{eq:XLfact}
    X_\pi = 1.00\, (1.00
    )\,, \quad     X_K = 1.02\, (1.01
    )\,, \quad     X^*_\pi = 0.99\, (1.00
    )\,, \quad     X^*_K = 0.93 \,(0.95
    )\,.
\end{equation}
For comparison, we give in brackets numerical results obtained with the full expression in \eqref{eq:XLFull} using the QCD sum-rule results from \cite{Gubernari:2018wyi}. We note that for $B\to D^*$ decays the corrections can be at the percent-level. Due to the excellent agreement between the two approaches, for our numerical analysis we take the first number as the central value and add the difference as an additional uncertainty. The uncertainty thus obtained is in fact the same as the uncertainty on $X_L$ obtained using the form factors from \cite{Gubernari:2018wyi}. 

Finally, combining the NNLO QCD value for $a_1$ listed in Table~\ref{tab:a1vals}, the $X_L$ factors in \eqref{eq:XLfact}, our QED results in \eqref{eq:deltaQED}$-$\eqref{eq:deltaQEDlast} and $\delta_{\rm U}$ (in brackets $\delta_{\rm U}^{(0)}$), we obtain the new predictions for $R_L^{(*)}$ listed in Table~\ref{tab:nonlep2semilep}. For the QCD NNLO results, we include uncertainties coming from $a_1^{\rm QCD}$ as before and those from CKM factors and $f_L$. The final uncertainty is dominated by the uncertainty on $a_1^{\rm QCD}$. For the QED contribution, the quoted uncertainty is the QCD NNLO uncertainty with the uncertainty from $U(D^{(*)}L)$ given in \eqref{eq:usoftval} added in quadrature. This uncertainty is rather conservative and therefore we do not include other uncertainties for the QED contributions. 

 \vspace*{0.16cm}
 \begin{table}
 \begin{center}
 \begin{tabular}{lcccc c}
 \hline \hline
 &&&&\\[-0.5cm]
 $R_L^{(*)}$ & ${\rm LO}$ & ${\rm QCD~NNLO}$ & $+\delta_{\rm QED}$ & $+\delta_{\rm U}\, (\delta_{\rm U}^{(0)})$  \\
 \hline
 &&&&\\[-0.3cm]
   $R_\pi$
 & $\phantom{-} 0.969 \pm 0.021$
 & $\phantom{-} 1.078^{+0.045}_{-0.042}$
 & $\phantom{-} 1.069^{+0.045}_{-0.041}$
 & $\phantom{-} 1.074^{+0.046}_{-0.043}\,(1.003^{+0.042}_{-0.039})$\\ \addlinespace
   $R_\pi^*$
 & $\phantom{-} 0.962 \pm 0.021$
 & $\phantom{-} 1.069^{+0.045}_{-0.041}$
 & $\phantom{-} 1.059^{+0.045}_{-0.041}$
  & $\phantom{-} 1.065^{+0.047}_{-0.042}\,(0.996^{+0.043}_{-0.039})$\\ \addlinespace
 \hline
 &&&&\\[-0.3cm]
   $R_K \cdot 10^2$
 & $\phantom{-} 7.47 \pm 0.07$
 & $\phantom{-} 8.28^{+0.27}_{-0.26}$
 & $\phantom{-} 8.21^{+0.27}_{-0.26}$
  & $\phantom{-} 8.44^{+0.29}_{-0.28}\,(7.88^{+0.26}_{-0.25})$\\ \addlinespace
   $R_K^* \cdot 10^2$
 & $\phantom{-} 6.81 \pm 0.16$
 & $\phantom{-} 7.54^{+0.31}_{-0.29}$
 & $\phantom{-} 7.47^{+0.30}_{-0.29}$ 
 & $\phantom{-} 7.68^{+0.32}_{-0.30}\,(7.19^{+0.29}_{-0.28})$  \\ \addlinespace
 \hline \hline

 \vspace*{-32pt}

 \end{tabular}
 \end{center}
 \caption{\label{tab:nonlep2semilep} Theoretical predictions for $R_L^{(*)}$ expressed in $\text{GeV}^2$ at LO, NNLO QCD and subsequently adding $\delta_{\rm QED}$ given in \eqref{eq:deltaQED} and the ultrasoft effects $\delta_{\rm U}$ (or in brackets $\delta_{\rm U}^{(0)}$). The last column presents our final results.}
\end{table}


\section{Discussion}

We discussed the QCD$\times$QED factorization of $B$ meson decays into heavy-light final states. Compared to the light-light final states \cite{Beneke:2020vnb}, the factorization is simpler as spectator scattering does not play a role, but the hard-scattering kernels are more complicated due to an additional dependence on the heavy charm-quark mass. After discussing the generalized factorization formula, we presented new predictions for the tree-level coefficient $a_1$ including NNLO QCD and the leading $\mathcal{O}(\alem)$ QED corrections. We find sub-percent QED effects coming from the hard-scattering kernels and Wilson coefficients, at the same level as the QCD uncertainties. On the other hand, for the decay rate, ultrasoft effects play a dominant role which can be several percent.

In Table~\ref{tab:nonlep2semilep}, we presented our results for observable ratios $R_L$ between the non-leptonic and semi-leptonic rates. The last column of Table~\ref{tab:nonlep2semilep} shows our final results which consists of the NNLO QCD predictions including QED effects for the $R_L(\Delta E)$ as defined in \eqref{eq:nonlep2semilepII} (for $R_L^{(0)}(\Delta E)$ as defined \eqref{eq:nonlep2semilep}). We stress the importance of the consistent handling of the ultrasoft effects between theory and experiment; the values and those in brackets differ significantly. Therefore, we do not give the current experimental values for the ratios which depend on the treatment of ultrasoft effects in the semi-leptonic rate (and in fact also of that in the non-leptonic rate). Preferably either the non-radiative or the soft-photon inclusive semi-leptonic differential decay would be given by the experimental collaborations directly.

Finally, we note that we did not present updated results of the branching ratios, as those would require inputs on the form factors for which QED effects are currently not known. This problem can be circumvented using the ratio $R_L$. However, in principle, our results for $a_1$ in Table~\ref{tab:a1vals} plus the ultrasoft effects in \eqref{eq:usoftval} provide predictions for the branching ratios employing a specific form factor parametrization with a possible additional uncertainty for the neglected QED form factor effects. Even in that case, the QED corrections would be dominated by ultrasoft effects. Comparing the theory predictions with experimental data then requires a careful study to ensure that the same observable is compared. Experimentally, QED effects are usually implemented using the PHOTOS Monte Carlo \cite{Golonka:2005pn}, and have been studied in a multitude of $B$ decays (see e.g. \cite{Carbone:2010wza, Cali:2019nwp,Bordone:2016gaq}). However, in light of the recent attention given to $B\to D L$ decays and the discrepancies that were observed between the experimental data and the theoretical predictions for the branching ratios \cite{Bordone:2020gao, Cai:2021mlt}, a renewed and more detailed study of these ultrasoft effects seems timely.

\subsubsection*{Acknowledgements}
We thank Tobias Huber and Alex Khodjamirian for discussions. This research was supported in part by the Deutsche
Forschungsgemeinschaft (DFG, German Research Foundation) through
the Sino-German Collaborative Research Center TRR110 “Symmetries
and the Emergence of Structure in QCD” (DFG Project-ID 196253076, NSFC Grant No. 12070131001, - TRR 110)

\appendix
\section{Convoluted kernels}
\label{sec:app:expr}

The convoluted kernels are defined as
\begin{equation}
\int_0^1 du \,T^{(*)(1)}(u,z)\, \phi_L(u) = \sum_{k=0}^2 \alpha_k^L(\mu)\biggl[V^{(1)}_{k-}(z) \pm \sqrt{z} V^{(1)}_{k+}(z) \biggr] \equiv \sum_{k=0}^2 \alpha_k^L(\mu) V_k^{(*)(1)}(z)  \,,
\end{equation}
where
\begin{equation}
\begin{split}
V_{0-}^{(1)}(z) &= -\frac{5 L_b}{3}-\frac{2 L_\nu}{3}-\frac{2 (2 z^3-6 z^2-6 z+1)}{3 (z-1)^3}\text{Li}_2(z) \\
&-\frac{2 (z-3) z^2 }{3 (z-1)^3}\ln^2 z +\frac{4 (z^3+(-3-i \pi ) z^2+3 i \pi  z+z+1) z }{3 (z-1)^3}\ln z \\
&+\frac{-4 z^4+15 z^3-2 z^2+8 z+1}{3 (z-1)^2 z}\ln (1-z) \\
&+ \frac{24 i \pi  z^4+4 (-19-29 i \pi +\pi ^2) z^3+3 (45+52 i \pi -4 \pi ^2) z^2}{18 (z-1)^3} \\
&+ \frac{-6 (25+18 i \pi +2 \pi ^2) z+2 \pi ^2+44 i \pi +91}{18 (z-1)^3} \,,
\end{split}
\end{equation}

\begin{equation}
\begin{split}
V_{1-}^{(1)}(z) &= -\frac{L_b}{2}+\frac{2 z (8 z^2+12 z+5) }{(z-1)^4}\text{Li}_2(z) +\frac{2 z^2 (4 z+3) }{(z-1)^4}\ln ^2 z \\
& +\frac{2 z (-5 z^3+6 i (4 \pi +5 i) z^2+3 (11+6 i \pi ) z+2)}{3 (z-1)^4} \ln z \\
&+\frac{(23 z^3+145 z^2+127 z+5) }{6 (z-1)^3}\ln (1-z) \\
&-\frac{101 z^4+1688 z^3+54 z^2+24 \pi ^2 (8 z^2+12 z+5) z}{72 (z-1)^4} \\
&-\frac{24 i \pi  (11 z^4+56 z^3-60 z^2-8 z+1)-1976 z+133}{72 (z-1)^4} \,,
\end{split}
\end{equation}

\begin{equation}
\begin{split}
V_{2-}^{(1)}(z)&= -\frac{3 L_b}{5}+\frac{12 z (4 z^2+6 z+1)}{(z-1)^5}  \text{Li}_2(z)+\frac{12 z^2 (z+1) }{(z-1)^5}\ln ^2 z \\
&-\frac{4 z (z^3+(9-6 i \pi ) z^2+(-9-6 i \pi ) z-1) }{(z-1)^5}\ln z \\
&+\frac{(3 z^4+23 z^3+463 z^2+163 z+8) }{5 (z-1)^4}\ln (1-z) \\
&+\frac{-861 z^5+1580 z^4-33610 z^3+27060 z^2-600 \pi ^2 (4 z^2+6 z+1) z}{300 (z-1)^5} \\
&+\frac{-120 i \pi  (z^5+5 z^4+100 z^3-100 z^2-5 z-1)+5095 z+736}{300 (z-1)^5} \,,
\end{split}
\end{equation}

\begin{equation}
\begin{split}
V_{0+}^{(1)}(z) &= \frac{2 (4 z^2+10 z+1)}{3 (z-1)^3} \text{Li}_2(z) +\frac{4 z (z+2)}{3 (z-1)^3}\ln^2 z \\
&+\frac{4 ((-5+2 i \pi ) z^2+(4+4 i \pi ) z+1)}{3 (z-1)^3}\ln z +\frac{(20 z^3+12 z^2-3 z+1) }{3 (z-1)^2 z^2}\ln (1-z) \\
&+\frac{-(93+120 i \pi +8 \pi ^2) z^3+4 \pi  (-5 \pi +24 i) z^2+(99+24 i \pi -2 \pi ^2) z-6}{18 (z-1)^3 z} \,,
\end{split}
\end{equation}

\begin{equation}
\begin{split}
V_{1+}^{(1)}(z) &= -\frac{2 (4 z^3+22 z^2+13 z+1)}{(z-1)^4}\text{Li}_2(z) -\frac{4 z (z^2+5 z+2)}{(z-1)^4}\ln^2 z \\
&+\frac{4 ((19-6 i \pi ) z^3+(9-30 i \pi ) z^2+(-27-12 i \pi ) z-1)}{3 (z-1)^4} \ln z \\
&-\frac{77 z^2+140 z+23}{3 (z-1)^3} \ln (1-z) +\frac{259 z^3+771 z^2+6 \pi ^2 (4 z^3+22 z^2+13 z+1)}{18 (z-1)^4} \\
&+\frac{24 i \pi  (19 z^3+9 z^2-27 z-1)-879 z-151}{18 (z-1)^4} \,,
\end{split}
\end{equation}

\begin{equation}
\begin{split}
V_{2+}^{(1)}(z) &= \frac{4 (4 z^4+42 z^3+60 z^2+18 z+1)}{(z-1)^5}\text{Li}_2(z) +\frac{8 z (z^3+10 z^2+12 z+2)}{(z-1)^5}\ln^2 z \\
&+\frac{4 ((-45+12 i \pi ) z^4+8 i (15 \pi +17 i) z^3+36 (3+4 i \pi ) z^2+24 (3+i \pi ) z+1)}{3 (z-1)^5}\ln z \\
&+\frac{(181 z^3+797 z^2+473 z+49) }{3 (z-1)^4}\ln (1-z) +\frac{-1003 z^4-8068 z^3+1602 z^2}{36 (z-1)^5} \\
&+\frac{-2 \pi ^2 (4 z^4+42 z^3+60 z^2+18 z+1)-4 i \pi  (45 z^4+136 z^3-108 z^2-72 z-1)}{3 (z-1)^5} \\
&+\frac{7012 z+457}{36 (z-1)^5} \,.
\end{split}
\end{equation}

\bibliographystyle{JHEP} 
\bibliography{refs.bib}

\end{document}